\documentclass[11pt, a4]{article}

\usepackage{amsmath,amssymb,graphicx}
\usepackage{lineno}
\usepackage[usenames,dvipsnames]{color}
\usepackage{subfigure}
\usepackage[margin=1in]{geometry}
\usepackage{yhmath}
\usepackage{ulem}
\usepackage{appendix}

\usepackage{cancel}

\newcommand{\al}{\alpha}

\def\be{\begin{equation}}
\def\ee{\end{equation}}
\newcommand{\bs}{\boldsymbol}

\usepackage[colorlinks=true, pdfstartview=FitV, linkcolor=darkblue, citecolor=darkblue, urlcolor=darkblue]{hyperref}
\definecolor{darkblue}{rgb}{0.1,0.1,0.45}

\def\C{{\mathbb C}}

\def\R{{\mathbb R}}

\def\T{{\mathbb T}}
\def\Z{{\mathbb Z}}

\def\Rscr{\mathcal R}

\usepackage{verbatim}
\def \eqref#1{(\ref{#1})}

\makeatletter
\@addtoreset{equation}{section}
\makeatother

\def\le{\left}
\def\ri{\right}

\def\bc{\begin{corollary}}
\def\ec{\end{corollary}}
\def\&{&{\hskip -20pt}}

\def \s{\mathfrak s}

\def\br{\begin{remark}\rm\small}
\def\1{{\bf 1}}
\def\er{\end{remark}}
\def\bt{\begin{theorem}}
\def\et{\end{theorem}}

\def\bx{\begin{example}}
\def\ex{\end{example}}
\def\bi{\begin{itemize}}
\def\ei{\end{itemize}}
\def\bd{\begin{definition}}
\def\ed{\end{definition}}
\def\bp{\begin{proposition}\rm}
\def\bl{\begin{lemma}\em}
\def\el{\end{lemma}}
\def\ep{\end{proposition}}
\def\bea{\begin{eqnarray}}
\def\eea{\end{eqnarray}}

\def\C{{\mathbb C}}
\def\R{{\mathbb R}}

\def\Z{{\mathbb Z}}

\newtheorem{theorem}{Theorem}[section]

\newtheorem{example}[theorem]{Example}
\newtheorem{coroll}[theorem]{Corollary}
\newtheorem{examps}[theorem]{Examples}

\newtheorem{lemma}[theorem]{Lemma}
\newtheorem{remark}[theorem]{Remark}
\newtheorem{remarks}{Remarks}
\newtheorem{proposition}[theorem]{Proposition}
\newtheorem{definition}[theorem]{Definition}

\def\br{\begin{remark}}
\def\er{\end{remark}}
\def\bt{\begin{theorem}}
\def\u{\mathbf u}

\def\et{\end{theorem}}
\def\bc{\begin{coroll}}
\def\ec{\end{coroll}}
\def\brs{\begin{remarks} \rm\
\begin{enumerate}}
\def\ers{\end{enumerate}\end{remarks}}
\def\bl{\begin{lemma}}
\def\el{\end{lemma}}
\def\bxs{\begin{examps}. \rm\begin{enumerate}}
\def\exs{\end{enumerate}\end{examps}}
\def\bd{\begin{definition}}
\def\ed{\end{definition}}
\def\bp{\begin{proposition}}
\def\ep{\end{proposition}}
\def\be{\begin{equation}}
\def\ee{\end{equation}}

\def\bea{\begin{eqnarray}}
\def\eea{\end{eqnarray}}
\def\beas{\begin{eqnarray*}}
\def\eeas{\end{eqnarray*}}
\def\gt{\hat\gamma}
\def \hf{\frac{1}{2}}

\def\part{\partial}

\def \ra{\rightarrow}

\def\C{{\mathbb C}}

\def \Th{\Theta}

\def\a{\alpha}
\def\b{\beta}
\def\d{\delta}
\def\g{\gamma}
\def\k{\varkappa}

\def\o{\omega}
\def\r{\rho}
\def\s{\sigma}

\def\e{\varepsilon}
\def\z{\zeta}

\def\R{{\mathbb R}}

\def\Z{{\mathbb Z}}

\date{}

\begin{document}

\title{\bf Rogue waves in multiphase solutions  of the focusing NLS equation}

\author{M.~Bertola$^{1,2}$, G.~A.~El$^3$, A.~Tovbis$^4$ \\ \\
$^1$ Department of Mathematics, Concordia University, Canada \\
$^2$ Area of Mathematics, SISSA/ISAS, Trieste, Italy \\
$^3$ Department of Mathematical Sciences, Loughborough University, UK \\
$^4$  Department of Mathematics, University of Central Florida, USA
}

\maketitle
\
\begin{abstract}
Rogue waves appearing on deep water or in optical fibres are often modelled by certain breather solutions of the focusing nonlinear Schr\"odinger (fNLS) equation which are referred to as solitons on finite  background (SFBs). 
A more general modelling of  rogue waves can be achieved via the consideration of  multiphase, or  finite-band, fNLS solutions of whom the standard  SFBs and the structures forming due to their collisions represent particular, degenerate, cases.
A generalised rogue wave notion then naturally enters as a large-amplitude localised coherent structure occurring within a finite-band fNLS
solution. In this paper, we use the winding of real tori to show the mechanism of the appearance of such generalized rogue waves and derive an analytical criterion distinguishing finite-band potentials of the fNLS equation that exhibit  generalised rogue waves.
\end{abstract}


\maketitle

\section{Introduction}
Rogue waves are waves of unusually large amplitude  $A_m$, whose appearance statistics deviates from the Gaussian distribution by exhibiting ``heavy tails'' in the probability density function (see  \cite{suret2015} and references therein).  The conventional amplitude criterion for rogue waves  is  
$A_m/A_s > 2$, where $A_s$ is the significant wave height  defined in oceanography as the average wave height (trough to crest) of the highest third of waves  (see e.g. \cite{kharif_book},  \cite{agafontsev_zakharov}).  
To have some workable amplitude criterion one can use the significant wave height computed over  Gaussian statistics. For a random complex Gaussian wave field $\psi(x,t)$  one has the Rayleigh probability distribution function for  $|\psi|$ (see e.g. \cite{nazarenko}), resulting in the convenient formal amplitude criterion for rogue waves:  $A_m^2 / \langle |\psi|^2 \rangle  >8$,   where $\langle |\psi^2| \rangle$ is the mean value of the wave intensity (see e.g. \cite{agafontsev_zakharov}).

 Recent experiments in water waves \cite{onorato_2004}, \cite{chabchoub} and in fibre optics \cite{rogue_nat2007}, \cite{kibler_peregrine_2010}, \cite{kibler_km}, \cite{kibler_ab} (see also review articles \cite{turitsyn2013}, \cite{dudley_nat2014}) have convincingly demonstrated that rogue waves are a generic physical phenomenon deserving a comprehensive investigation \cite{onorato_etal2013}, \cite{turitsyn2013}.
  It has also been understood that rogue waves play important  role in the characterization of the nonlinear stage of modulational instability \cite{zakharov_gelash2013}, \cite{zakharov_gelash2014} and particularly, in the development of integrable turbulence \cite{zakharov2009}, \cite{agafontsev_zakharov}, \cite{suret2015}.  One of the universal mathematical models for the rogue wave description is the one-dimensional focusing Nonlinear Schr\"odinger (fNLS) equation,
\be  \label{NLS}
i  \psi_t +  \psi_{xx} +2 |\psi|^2 \psi=0. 
\ee
The fNLS equation has a number of relatively simple exact solutions that  are often considered as analytical ``prototypes'' or models of rogue waves, the principal representatives being the Akhmediev breather (AB), the Kuznetsov-Ma (KM) breather and the Peregrine breather (see e.g. \cite{dysthe99}). These solutions represent {\it solitons on finite background} (SFBs), and their amplitude can satisfy the described above formal rogue wave criterion, the background amplitude $A_0$ being the amplitude of the underlying plane wave.  All three above types of SFBs have been realised in physical experiments (see \cite{turitsyn2013} and references therein). In these experiments,  initial conditions have been carefully designed to instigate the occurrence of the particular type of a rogue wave.  

Physical mechanisms of the ``spontaneous'' generation of rogue waves  have been the subject of many research studies (see e.g. 
\cite{onorato_etal2013}, \cite{turitsyn2013},  \cite{dudley_nat2014} and references therein). Some  of them relate the rogue wave appearance to the development of modulational instability of the plane wave  due to small perturbations (see, e.g., \cite{shrira}, \cite{agafontsev_zakharov}, \cite{zakharov_gelash2014}) or  large-scale initial modulations \cite{grim_tovbis}.  Other proposed mechanisms  involve  interactions of individual solitons \cite{dudley2010}, \cite{akhmediev_2016} or interaction of solitons with the plane wave \cite{zakharov_gelash2013}.
The appearance of  the higher-order rogue waves has been attributed to the interactions of elementary SFBs  \cite{akhmediev_PLA2009}.   In all cases,  the  modelling of individual rogue waves has been done within the framework of the solitary wave structures: either $N$-solitons or SFBs.  
However, recent analytical \cite{BT1}, \cite{ekt2015}  and numerical \cite{randoux_periodic2015} studies of the large-amplitude wave generation from  rather general classes of initial data  strongly suggest that typical rogue waves are generally described by the so-called {\it finite-band}, or multiphase, NLS solutions (also often referred to as finite-gap solutions) \cite{belokolos_etal_1994}, \cite{tracy_chen} of whom $N$-solitons, SFBs and the structures forming due to their collisions represent special degenerate cases, see \cite{osborne_book}. 

Finite-band solutions appear as a leading order approximation of the evolving fNLS solution (and thus,  enter the rogue wave theory)  via two basic scenarios realised within the small-dispersion (semi-classical) fNLS evolution framework. In the first scenario,  finite-band potentials exhibiting high local amplitudes approximate the coherent structures regularising the gradient catastrophe forming in the evolution of an (analytic)   modulated high frequency plane wave solution \cite{BT1}. The prominent feature of this scenario is the formation of expanding chains of Peregrine breathers right beyond the gradient catastrophe point. In the second scenario, the high-amplitude breather lattices are formed as a result of interaction of dispersive shock waves (dam break flows) generated in the fNLS evolution of a rectangular initial potential \cite{ekt2015}.

Motivated by the previous studies \cite{osborne_book}, \cite{ekt2015}, \cite{randoux_periodic2015}, we  introduce  in this paper    the notion of a generalised NLS rogue wave as a large-amplitude localized ``fluctuation'' appearing within a generic finite-band NLS solution. Given some formal amplitude criterion for such a rogue wave event, it is clear that not all finite-band solutions of the fNLS equation can exhibit rogue waves.
By employing the recently obtained (\cite{BT}) explicit formula for the maximum of the  wave field amplitude for  finite-band potentials, see \eqref{max},
 (for the genus 2 case this result was established in \cite{wright}),
 we provide a simple analytical criterion for distinguishing the finite-band potentials  exhibiting (generalised) rogue waves. 

Finite-band potentials are known to be quasiperiodic functions of $x$ and $t$ \cite{belokolos_etal_1994} specifying integrable dynamics on multi-dimensional tori, the dimension of the torus being equal to the genus  of the  hyperelliptic Riemann surface on which the finite-band potential is defined.  We then use the winding of real tori to explain the occurrence of rogue waves within finite-band potentials.  The natural (uniform) probability measure on the torus $\mathbb{T}^g$ gives rise to a random process generated by the function $\psi_g(x,t)$ (see  \cite{ekmv2001} for the corresponding construction in the framework of the Korteweg -- de Vries equation). This  makes possible the determination of the statistics of generalised rogue waves occurring within finite-band potentials. Such a statistical study  will be the subject of a separate work.

\section{Rogue waves on a finite-band potential background}

The simplest solution to the fNLS equation (\ref{NLS}) is a plane wave (sometimes called a ``condensate''),
 \begin{equation}\label{pw}
\psi= q e^{2iq^2  t} \equiv \psi_0.
\end{equation}
The plane wave solution (\ref{pw}) is well known to be modulationally unstable with respect to small long-wave perturbations \cite{tracy_chen}.
As is widely appreciated (see e.g. \cite{tracy_chen}, \cite{osborne_book}, \cite{flee86}), one of the  general  mathematical frameworks for the description of the development of modulational instability is the so-called finite-gap theory \cite{belokolos_etal_1994}, which is a (nontrivial) extension of the IST to  fNLS with periodic boundary conditions \cite{IK},  \cite{ablowitz_ma}.  
A $g$-phase 
 finite-band solution of (\ref{NLS}) is defined in terms of the
Riemann theta-function  $\Theta_g$ associated with the hyperelliptic Schwarz symmetrical Riemann surface $\Rscr$
of genus $g$ specified by
\begin{equation}
\label{rsurf}
R(z)=\prod_{j=0}^g(z-\a_j)^\hf(z-\bar\a_j)^\hf,\qquad
\alpha_j = a_j + i b_j, \  \ b_j>0,
\end{equation}
$z \in \C$ being the complex spectral parameter (see e.g. \cite{belokolos_etal_1994}).
It will be convenient for us to write this solution in the form (\cite{KMM}, \cite{TVZ1}, \cite{BT})
\begin{equation}
\label{fingap}
 \psi_{g}(x,t)=\frac{\Theta_g(2\u_\infty+{\bs \eta}(x,t))\Th_g(0)}{\Theta_g(2\u_\infty)\Th(\bs \eta(x,t))}\sum_{j=0}^g b_je^{2iG(x,t)},
\end{equation}
where the phase vector
\be\label{phases}
\bs \eta(x,t)=\bs k\, t+\bs \omega\,x+\bs \eta^0
\ee
has (real) components $\eta_j =k_j x + \omega_j t + \eta_j^0$, $j=1,\dots, g$, the constant vector $\u_\infty\in\C^g$ is the 
value of the Abel map  on $\Rscr$ (with the base point $\a_0$) evaluated at $\infty_+$ (on the main sheet)  and $G(x,t)$ is a real valued,
linear in $x,t$ function (which  does not affect $|\psi_{g}|$ and  plays no role in this paper).
In \eqref{fingap},  the wavenumber vector  $\bs k=(k_1,\dots,k_g)$ and the frequency vector $\bs \omega =(\o_1,\dots,\o_g)$ are defined in terms of the branch points $\alpha_j$ alone, and $\bs \eta^0$ is the initial phase  vector.
 (In more technical terms,  vectors $\bs k,\bs \omega$ are the vectors of $\mathbf B$-periods of the normalised
meromorphic differentials $dP,dQ$ of the second kind  on $\Rscr$, which have poles only  at $\infty_\pm$
and have the corresponding principal parts $\mp\frac{1}{\z^2}d\z$,  $\mp\frac{2}{\z^3}d\z$    respectively, $\z=\frac 1{z}$.)
Based on \eqref{fingap}, a remarkably simple formula \eqref{max} for $\max  |\psi_{g}(x,t)|$  was recently proved  in \cite{BT}.

The  plane wave  (\ref{pw}) itself represents a genus zero solution and 
lives on the Riemann surface  $\Rscr$  specified by Eq. (\ref{rsurf}) with $g=0$ and $\a_0=iq$, $q>0$, i.e.  $R (z) =  \sqrt{(z - iq)(z + i q)}$. Thus the spectral portrait of the plane wave is a vertical branch cut between the simple spectrum points $\a_0=iq$ and $\bar \a_0=-iq$.

For $g \ge 1$  the theta-solution (\ref{fingap}) is a (quasi-)periodic function of $x$ and $t$ depending on $g$ nontrivial oscillatory phases $\eta_j(x,t)$, so that
$|\psi_g(x,t)| = f_g(\bs{ \eta})$, where $\bs{\eta}=(\eta_1,\dots, \eta_g)\in\R^g$ and $f_g(\bs{ \eta})= f_g(\bs{ \eta}+\bs{e})$ 
for an  arbitrary $\bs{ e}\in\Z^g$. 
Thus  $|\psi_g|=f_g$ can be viewed as a (smooth) function on the $g$-dimensional real torus $\T^g$. We denote by 
$|\psi_g|_m$ the maximal value of $f_g$ over $\T^g$.

The wavenumbers $k_j$  and the frequencies $\omega_j$ can be calculated through the following expressions (\cite{flee86}, \cite{tracy_chen})  
\begin{equation}
\label{kj}
k_j=-2 i  \k_{j, 1} \, , \qquad \omega_j= - 2 i  \left[  \tfrac12  \sum \limits_{k=0}^{g} (\al_k + \bar \al_k ) \  \k_{j,1} + \k_{j, 2} \right], \quad j=1, \dots, g,
\end{equation}
where $\k_{j,k}=\k_{j,k}(\bs{\al}, \bs{\bar \al})$ are found from the system (see (\ref{norm1}))
\begin{equation}
\label{holonorm}
\sum \limits_{i=1}^{g}\k_{j,i} \oint_{\mathbb{A}_k} \frac{\z^{g-i}}{R (\z)} d\z = \delta_{jk} \, , \quad j,k = 1, \dots, g.
\end{equation}
Here $\delta_{ik}$ is the Kronecker symbol, $\bs{\al}=(\a_0, \dots,\a_g)$  and  $\mathbb{A}_k$ is a  negatively (clockwise) oriented loop around the  branch cut connecting $\bar \a_k$ and $\a_k$.  Loosely speaking, the genus $g$ solution can be viewed as a ``nonlinear superposition'' of $g+1$ nonlinear modes (including the trivial, plane wave mode).

As was mentioned in the Introduction, the fNLS solutions traditionally considered as ``analytical prototypes'' for  rogue waves are the Akhmediev breather (AB), the Kuznetsov-Ma (KM) breather and their limiting case, the Peregrine breather (see e.g. \cite{dysthe99}). These SFB solutions represent degenerate 
genus two solutions of the fNLS equation (as the branch points $\a_1$ and $\a_2$ merge together) and are
spectrally defined by a basic branch cut between the points $\a_0$ and $\bar \a_0$ of the simple spectrum and two complex conjugate double points $\a_* \equiv \a_1=\a_2$, $\bar \a_* \equiv \bar \a_1= \bar \a_2$. If the rogue wave is stationary the double points $ \a_*$ and  $\bar \a_*$  are located  on the imaginary axis. 
Let $\a_0=i q$, $\a_*= i b_*$, $b_*>0$. If $b_*>q$, the genus two solution becomes the KM breather $\psi_M$, a time-periodic, spatially localised solution with the asymptotic behaviour $\psi_M \to \psi_0$ as $x \to \pm \infty$;
if $b_*<q$, then the solution is the spatially-periodic, time-localised AB  $\psi_{AB}$, so that $\psi_{AB} \to \psi_0$ as $t \to \pm \infty$ \cite{dysthe99}. Finally, if $b_*=q$ then the solution represents the Peregrine breather  $\psi_P$, which is localised both in time and space and has the asymptotic behaviour $\psi_P \to \psi_0$ as $(x,t) \to (\pm \infty, \pm \infty)$ \cite{dysthe99}.  
All the described solutions represent  the first-order SFBs. More sophisticated, higher order SFBs, are possible (see e.g. \cite{akhmediev_PRE09}), which represent degenerate fNLS solutions with genus $g>2$.  

 For the Peregrine breather the ``rogue wave ratio'' $K= |\psi_P|_m^2/ \langle |\psi_0|^2\rangle =9 >8$ (see the Introduction), 
   i.e. it  satisfies the amplitude rogue wave criterion.
 Here and henceforth, $|\psi|_m$ denotes the maximal amplitude of $\psi$. 
   For the KM breather, $K>9$ so it can also be considered as a possible prototype for a rogue wave. For the Akhmediev breather to be  considered as a rogue wave  prototype one must have $8 < K <  9$. We stress that in a physical context, all statements about the rogue wave nature of a large-amplitude coherent structure  should be considered in conjunction with the statistics of its occurrence.
 
Clearly, SFBs represent very special solutions of the fNLS equation.  Indeed, as was suggested in \cite{ekt2015} and explicitly demonstrated in  \cite{randoux_periodic2015},  the evolution of generic (including random) initial conditions  leads to  the formation of complex coherent nonlinear wave structures that  are locally well  approximated by modulated finite-band solutions $\psi_g(x,t)$ defined by the spectral branch points $\a_j$, $j=0,1, \dots , g$ (\cite{belokolos_etal_1994}, \cite{osborne_book}). 
The genus $g$ of the approximating solution is generally different in different regions of $x,t$-plane. These finite-band solutions can  exhibit very significant amplitudes $|\psi_{ g}|_m$ satisfying the {\it generalized rogue wave amplitude criterion}
\be\label{rwc}
K_g=\frac {|\psi_g|^{2}_m}{\langle |\psi^{2}_g|\rangle} > C_r,
\ee
where   $ \langle |\psi_g^2| \rangle$  is  the mean value  of the wave intensity $|\psi_g(x,t)|^2=f^2_g(\bs \eta)$ over the phase torus $\T^g$ (assuming incommensurability of the wave numbers  $k_j$ and  of the frequencies $\omega_j$ defined by (\ref{kj}).) 
The constant $C_r$ should be appropriately defined using statistical properties of finite-band solutions of the fNLS equation. 
In the calculations of this paper, to be consistent with the traditional rogue wave amplitude criterion for  SFBs, we shall be simply assuming  that $C_r=8$ with the understanding that all the quantitative conclusions made could be readily adapted to an arbitrary positive value of $C_r$.

 We now note that,  assuming incommensurable wavenumbers $k_j$  and incommensurable frequencies $\omega_j$, the average over the torus is equivalent to the spatial and temporal averages (ergodicity).  One of the consequences of ergodicity is that the value $\langle |\psi_g^2| \rangle$ does not depend on time.  It is also assumed in (\ref{rwc}) that $\langle |\psi^{2}_g| \rangle$ 
 is separated from zero, so that the fundamental NLS  solitons living on a zero background are excluded from the definition of a rogue wave. Clearly, for SFBs one should have $ \langle |\psi_g^2| \rangle= |\psi_0^2| =q^2$ (see Appendix for the proof for $g=2$).

The exact formula  for the average value of the intensity $ \langle |\psi_g^2| \rangle $
is derived in the Appendix, see \eqref{psi-av-fin1}.  With the additional  but nonrestrictive  simplifying assumption   $\sum_{j=0}^g a_j=0$, 
see \eqref{rsurf},
this formula becomes
 \be\label{psi-av-fin-simA}
 \langle |\psi_g^2| \rangle = 2 \sum_{j=1}^g\k_{j,1}\oint_{\mathbb{A}_j}\frac{z^{g+1}dz}{R(z)}+\sum_{j=0}^g (b_j^2-a_j^2),
\ee
where the coefficients $\k_{j,1}$ are defined by (\ref{holonorm}). As expected (see Appendix for the proof), for the  two-phase ($g=2$) finite-band solution $\psi$, the quantity (\ref{psi-av-fin-simA})  approaches the square of the amplitude of  the background (the ``condensate'') of the corresponding SFB.

 For the value of $|\psi_g|_m$, which is the maximum  of $|\psi_g|$ over $\T^g$,  an elegant and intuitive formula
 \be \label{max}
 |\psi_g|_m = \sum \limits_{j=0}^g b_j \, 
 \ee
 was found in \cite{BT} using the $g$-band solution  in the form (\ref{fingap})  (for $g=2$ case this result was established in \cite{wright}). Importantly, 
  if the  wavenumbers $k_j$ are incommensurable, the value $|\psi_g|_m$ is the supremum of $|\psi_{g}|$ over  $x \in \mathbb{R}$ for any fixed $t$. To attain this value  at a particular point $x,t$, or, more generally, within a certain $x,t$-domain, one has to  make an appropriate choice of the initial phases 
  $\bs{\eta}^0 \in \mathbb{T}^g$, see \eqref{fingap}, \eqref{phases}.
  A similar statement is true for  incommensurable frequencies $\o_j$ and the maximal value of $|\psi_g|$ for any fixed $x$.
 Assuming random (uniform) phase distribution over the torus one naturally arrives at the problem of the statistical description of the rogue wave occurrence within a given family of finite-band potentials. 
 
 In view of \eqref{psi-av-fin-simA} and \eqref{max},  the finite-band potential defined by the spectral points $\a_0, \a_1, \dots , \a_g$ and c.c. will exhibit rogue waves if
 \be\label{rwc_g}
 K_g(\bs{\a}, \bs{\bar \a}) = \frac{(\sum \limits_{j=0}^g b_j )^2}{2 \sum_{j=1}^g\k_{j,1}\oint_{\mathbb{A}_j}\frac{z^{g+1}dz}{R(z)}+\sum_{j=0}^g (b_j^2-a_j^2)} > C_r.
 \ee
 As was mentioned above, in what follows we will assume $C_r=8$ for the sake of definiteness.
 It is not difficult to show (see Appendix for details) that for the Akhmediev, Kuznetsov-Ma and Peregrine breathers formula (\ref{rwc_g}) transforms into the original rogue wave criterion (\ref{rwc}) with $\langle |\psi^{2}| \rangle=q^{2}$ and $|\psi|_m=q+ 2b_*$ (the latter formula appears in \cite{osborne_book}).  An important consequence of formula (\ref{rwc_g}) is that, even if the maximum local value $|\psi_g|_m$ attainable by the finite-band potential, is large, it does not guarantee that there are rogue waves within the potential due to the (possibly large) value of the averaged background $\langle |\psi^{2}_g| \rangle$ in the denominator of Eq. (\ref{rwc_g}). We stress that 
criterion (\ref{rwc_g}) applies to the whole family of finite-band potentials associated with a given Riemann surface $\Rscr $ and parametrised by the initial phase vector ${\bs \eta}^0 \in \mathbb{T}^g$. 
 
 \section{Winding of real tori and rogue waves}

\begin{center}
\begin{figure}[ht]
 \includegraphics[height=2in]{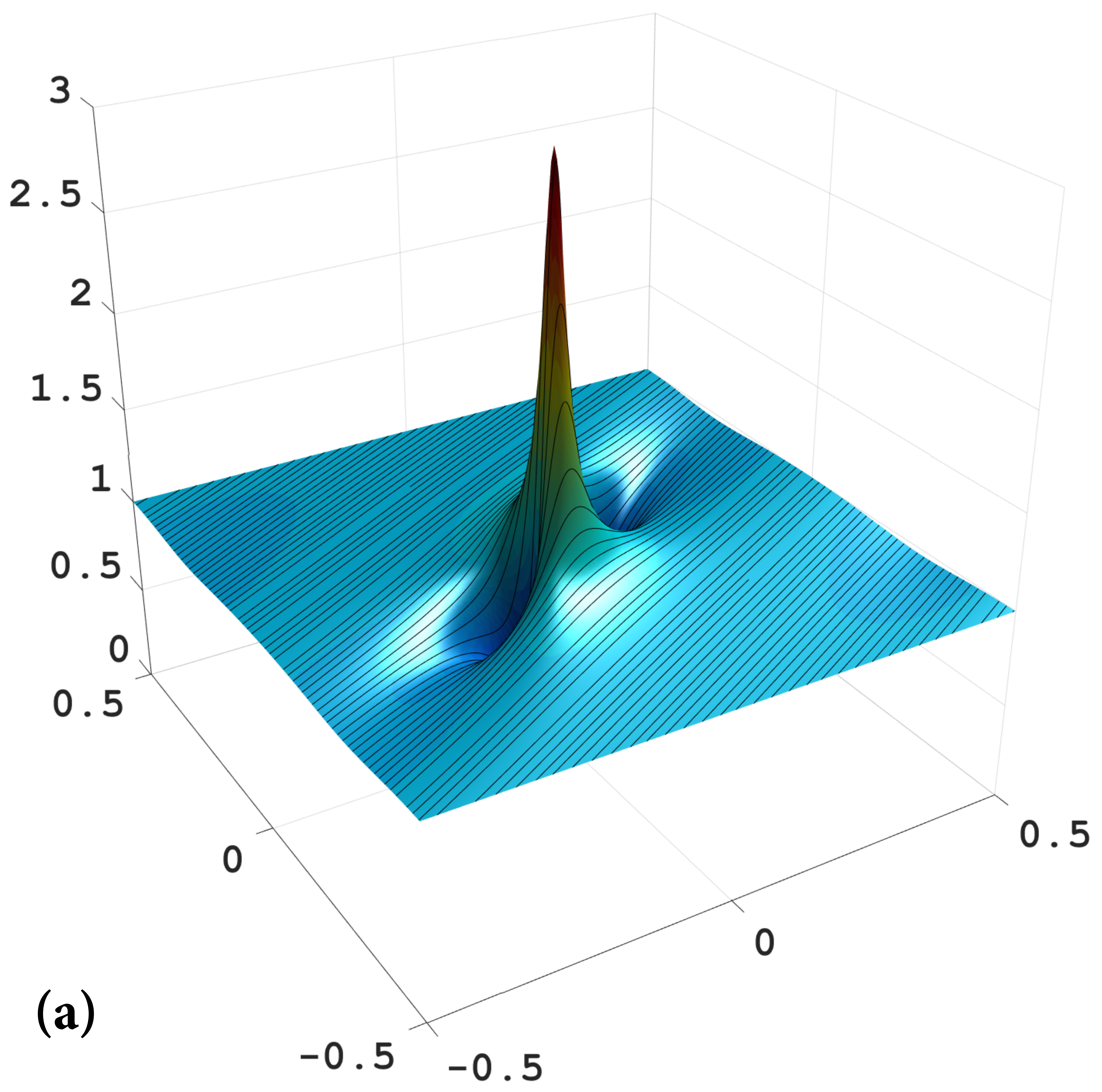}  \qquad
 \includegraphics[height=2in]{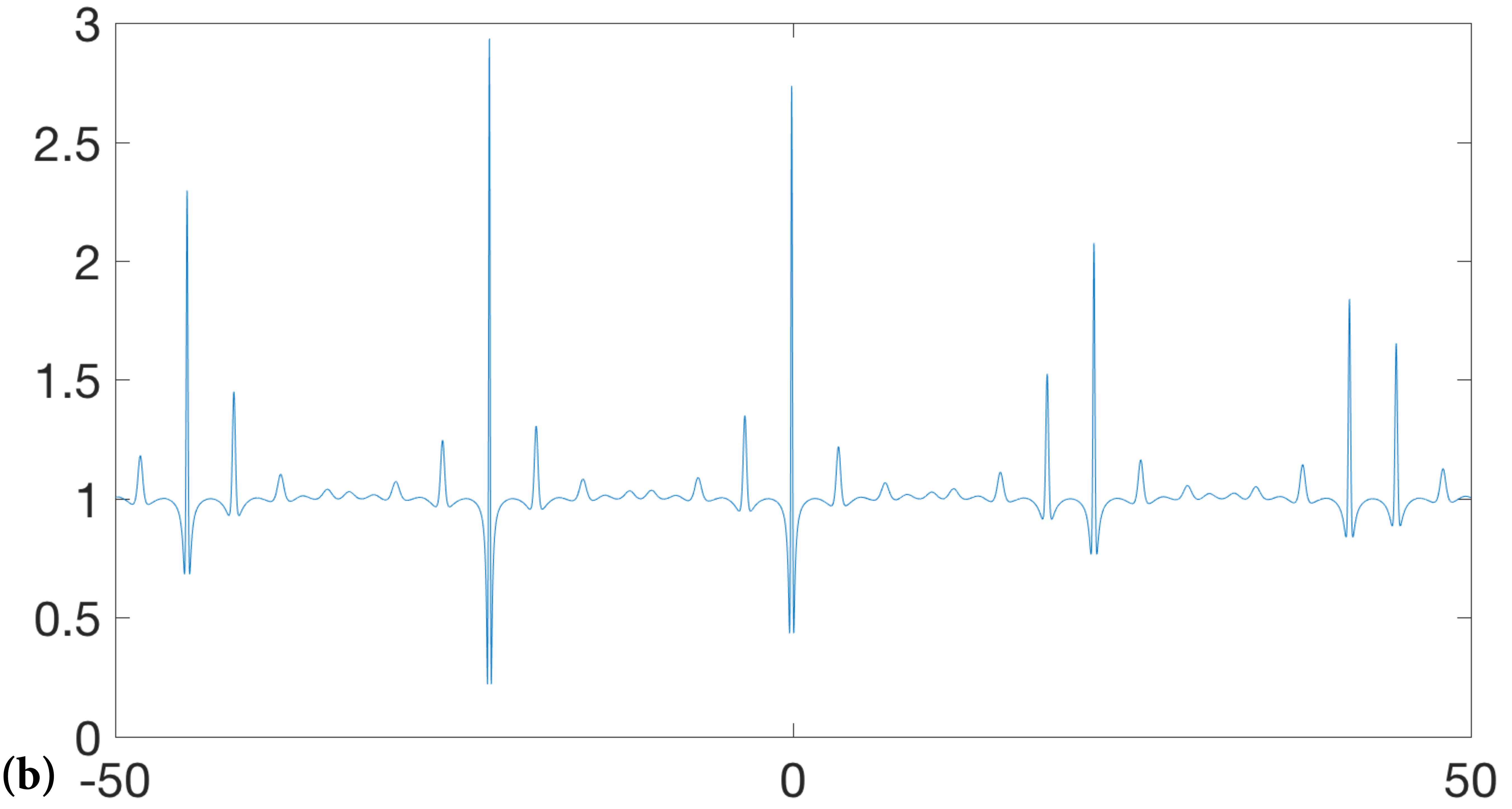} 
 \caption{$g=2$:  \ $\bs{\a}= (-0.0133+i,   -0.0033+i,    0.0167+i)$;
 ${\bf k} = (0.5830,    0.3147)$;
 $\langle |\psi_2^{{2}}| \rangle= 1.0266$; 
Max amplitude = 3; $K_2=8.76680$. 
(a) Winding of $\mathbb{T}^2$; \ (b) Plot of $|\psi_2(x)|$ corresponding to the winding in (a).
According to \eqref{rwc_g}, this configuration of the branchcuts 
 supports rogue waves. }
 \label{fig1}
 \end{figure}
\end{center}

\begin{center}
\begin{figure}[ht]
 \includegraphics[height=2in]{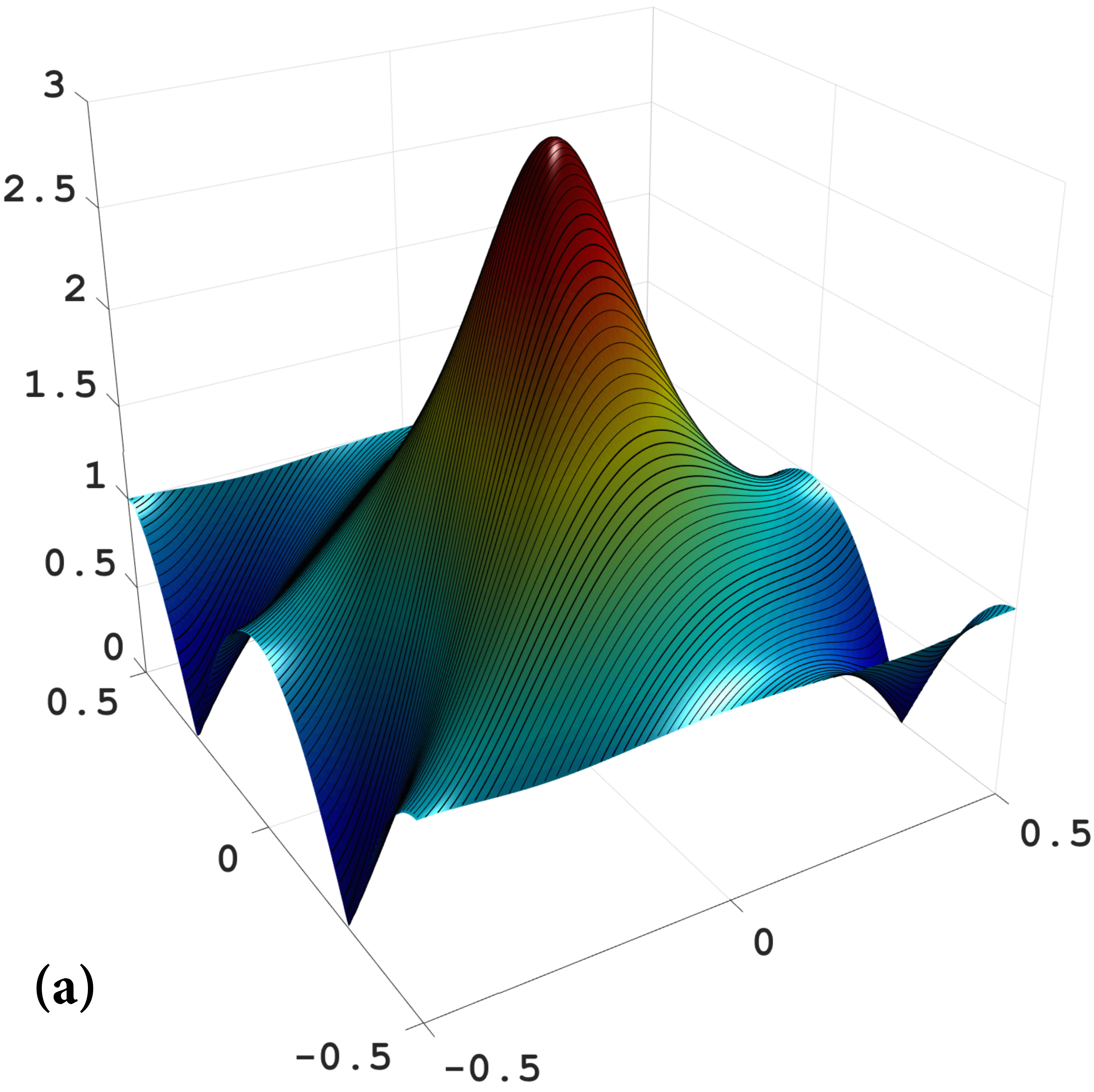}  
 \qquad 
 \includegraphics[height=2in]{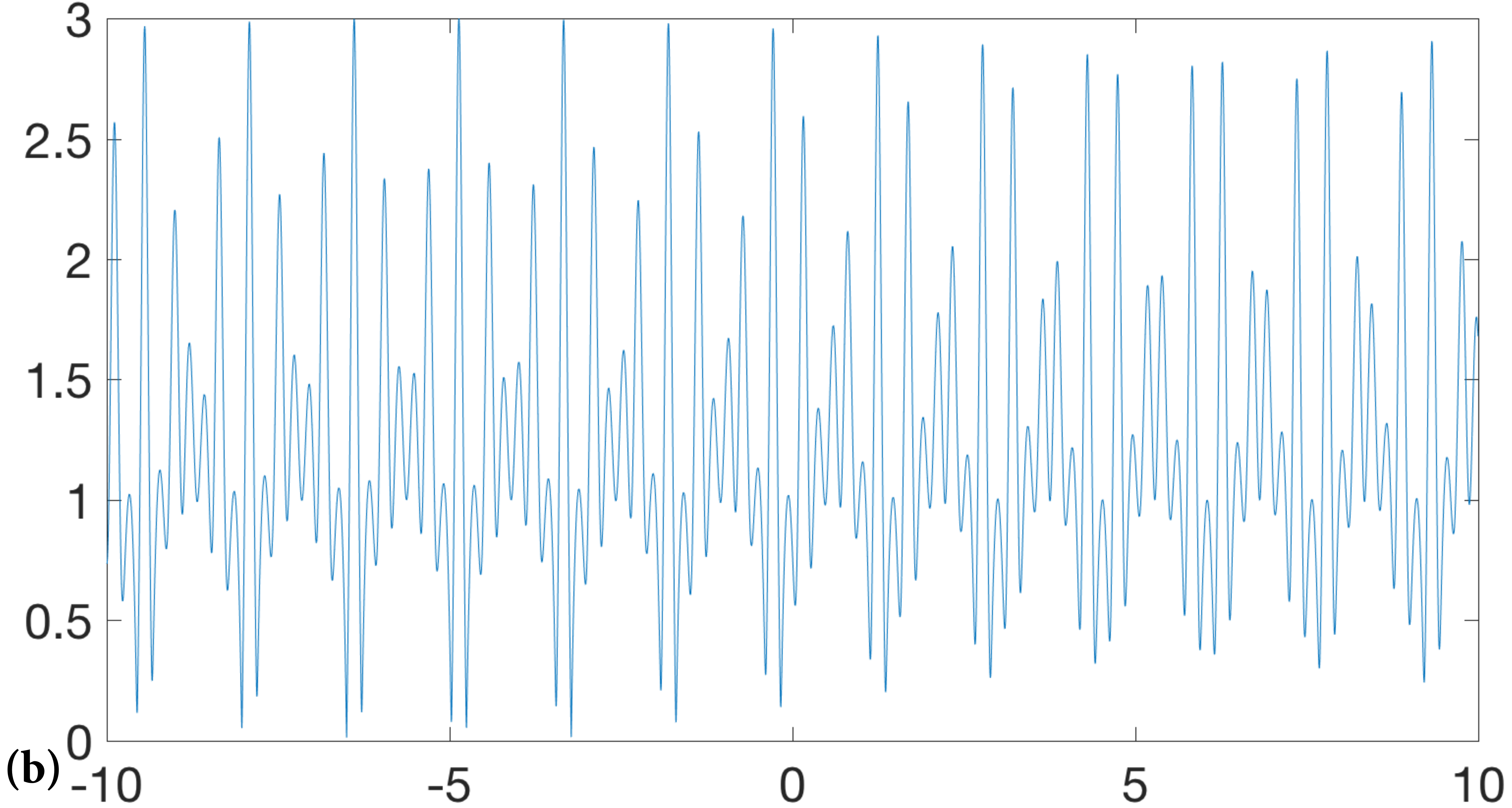}  
 \caption{$g=2$: \  $\bs{\a}=( -0.6667+i, -0.1667+i, 0.8333+i)$;
    ${\bf k} = (4.5811    2.6112)$;
 $\langle |\psi_2^{{2}}| \rangle= 1.9534$; 
Max amplitude $|\psi_2|_m= 3$; the rogue wave ratio $K_2=4.60735$.
(a) Winding of $\mathbb{T}^2$; \ (b) Plot of $|\psi_2(x)|$ corresponding to the winding in (a).
According to \eqref{rwc_g}, this configuration of the branchcuts 
does not support rogue waves.
}
 \label{fig2}
 \end{figure}
\end{center}

To get a better insight into the mechanism of the rogue wave occurrence within finite-band potentials we first consider the case $g=2$ which can be conveniently illustrated using the standard mapping of the two-dimensional torus $\T^2$ onto the square $D= [-0.5, 0.5]^2$ 
with coordinates $\eta_1,\eta_2$.
 Figs. \ref{fig1}a  and \ref{fig2}a show the graphs of $|\psi_2(\eta_1,\eta_2)|$ for ${(\eta_1, \eta_2)}  \in D$.  The values of the branchpoints of the spectral Riemann surface used for plotting Figs. \ref{fig1} , \ref{fig2} are chosen as follows. The plot in Fig. \ref{fig1} is constructed using formula (\ref{fingap}) with: 
$\bs{\a}= (-0.0133+i,   -0.0033+i,    0.0167+i)$ (case 1); Fig.~\ref{fig2}: $\bs{\a}=( -0.6667+i, -0.1667+i, 0.8333+i)$ (case 2).   
One can see that in both cases the  plots of $|\psi_2(x)|$ exhibit fairly regular structure of ``breather lattices''. As follows from (\ref{max}), in both cases  the amplitude maximum  $|\psi|_m=3$ but, due to the different values of the mean intensity $\langle |\psi_g^2|  \rangle$ the values of the rogue wave ratio 
 $K_2$ (\ref{rwc_g}) are different.   As a result, the potential in Fig. \ref{fig1} has $K_2= 8.77 >  8$ and hence, exhibits rogue waves while the one in Fig. \ref{fig2} has $K_2 =  4.61$ and is ``rogue wave free''. Notice that in the first case the three branchpoints $\a_j$, $j=0,1,2$ are close to each other
(and so are their c.c), so that the graph  of $|\psi_2|$ in Fig. \ref{fig1}a is close to the graph of the Peregrine breather. It is also evident that in this case the mean $\langle |\psi_2^{2}| \rangle$ is close to one. A slow evolution of a genus 2 breather lattice from a ``rogue wave free'' configuration to the configuration exhibiting rogue waves was shown   in \cite{ekt2015} to 
naturally occur in the semi-classical fNLS with initial data in the form of a rectangular barrier (the ``box'' problem).  
 
 The black curves in  Figs. \ref{fig1}a, \ref{fig2}a show the windings corresponding to the particular choices of the initial point $\bs \eta ^0$ on the torus. Each winding is parametrised by coordinate $x$, and its direction  is defined by the wavenumber vector ${\bf k}$ whose components $k_j$ are computed in terms of the branchpoints $\bs \a$  by formulae (\ref{kj}).  The dependencies $|\psi_2(x)|$ corresponding to the particular  windings in  Figs. \ref{fig1}a, \ref{fig2}a are shown in Figs. \ref{fig1}b and \ref{fig2}b. Given incommensurability of the values of $k_j$, each winding covers the corresponding phase torus densely  so  the probability of the rogue wave occurrence within the given potential $\psi_2(x,t; \bs{\a})$  is given by the ratio of the area of the part of the 2D torus confined to the level curve $|\psi_2(\eta_1, \eta_2)|=  C_r \langle |\psi_2| \rangle$ to the total area of the torus (the unit square).   
   
  \begin{figure}[ht]
\centerline{ \includegraphics[height=2.5in]{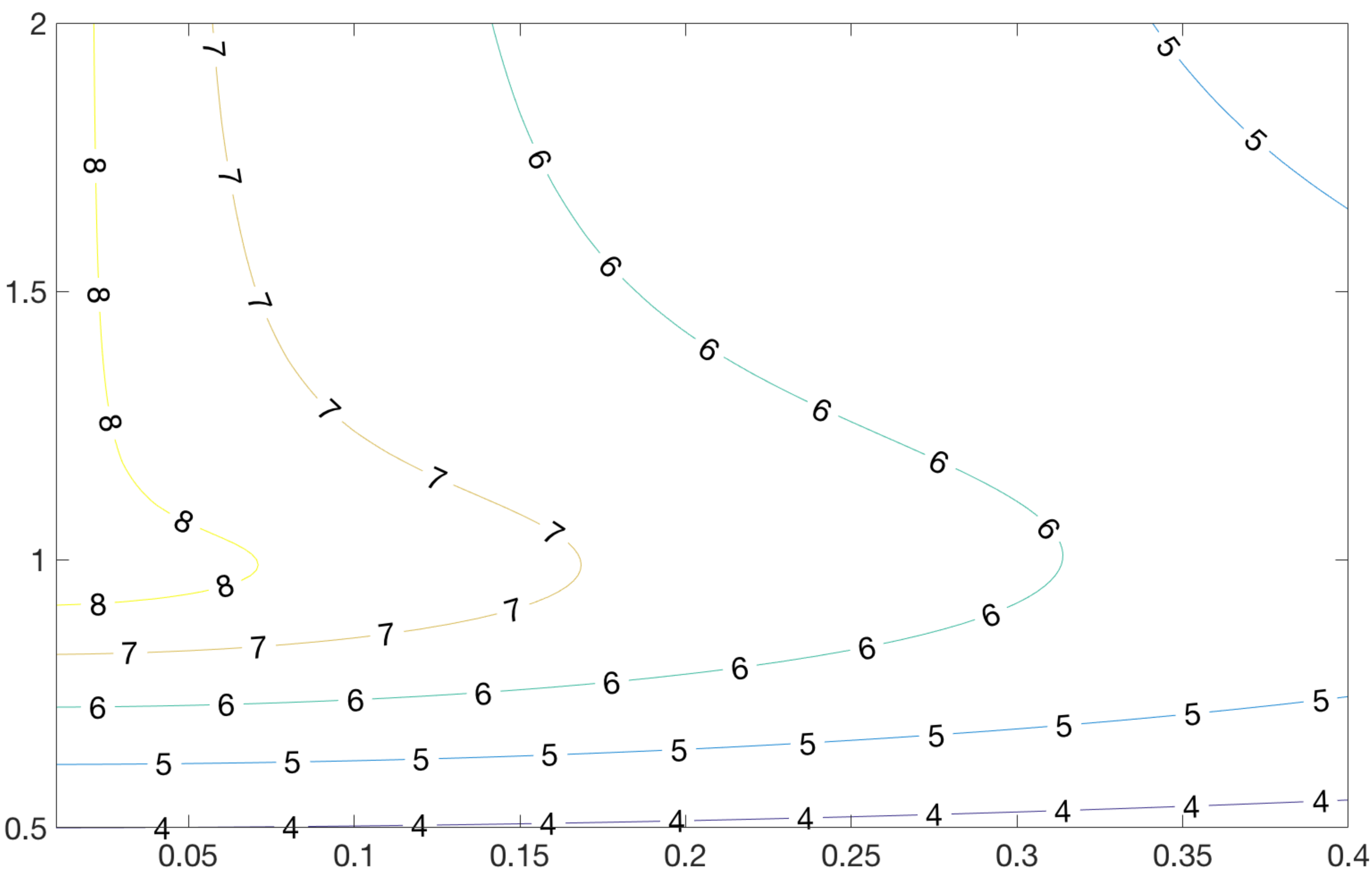} } 
 \caption{$g=2$. Level curves $K_2=4, 5, 6,7,8$  of the function  $K_2(\bs{\alpha},  \bs{\bar \a})$ (\ref{rwc_g}), where $\a_0=i$, $\a_1=-a+ib$, $\a_2=a+ib$, in the  $(a,b)$-plane. Only the  potentials with parameters within the area enclosed by the curve $K_2 = 8$ exhibit rogue waves.}
  \label{fig3}
 \end{figure}

 It is also instructive to plot the level curves of  $K_2({\bs{\a}, \bs{\bar\a}})$ (\ref{rwc_g}) in the space of the spectral parameters $\bs \a$. These  curves are shown on Fig. \ref{fig3} for a particular band configuration described below.
  Without loss of generality we  put $\a_0=i$, i.e. we set one of the bands to be located on the imaginary axis and to have the total height equals to $2$.  We shall call it the central band. We then place two other bands  of equal height  $b_1=b_2 =b$   symmetrically with respect to the central band  so that  $\a_1= - a+ib $, $\a_2= a+ib$. The level curves $b(a)$ of the function $K_2(\bs{\alpha},  \bs{\bar \a})$ (\ref{rwc_g}) for such potentials are shown in Fig.\ref{fig1} for $K_2=4,5,6,7,8$ (only the $a>0$ part of the countour plot is shown, the part with $a<0$ is symmetric with respect to the vertical axis). According to the criterion (\ref{rwc_g}) with $C_r=8$, rogue waves  are exhibited  only by the potentials with spectral parameters in the region enclosed by the curve $K_2=8$. One of the immediate conclusions one can make is that, for a genus 2 potential  with the described above symmetric band configuration to exhibit rogue waves, it has to have the side bands that: a) have sufficiently large height;  and  b) are located sufficiently close to the central band. The curve $K_2=8$ intersects the imaginary axis ($a=0$) at $b=\frac{\sqrt{8}-1}{2}$. One can see then that the  ABs with  $a=0$, $  \frac{\sqrt{8}-1}{2}  < b  < 1 $, the Peregrine breather ($a=0, b=1$) and the KM breathers ($a=0, b>1$) are all inside the ``rogue wave region'' as expected.

\begin{center}
\begin{figure}[ht]
 \includegraphics[height=2in]{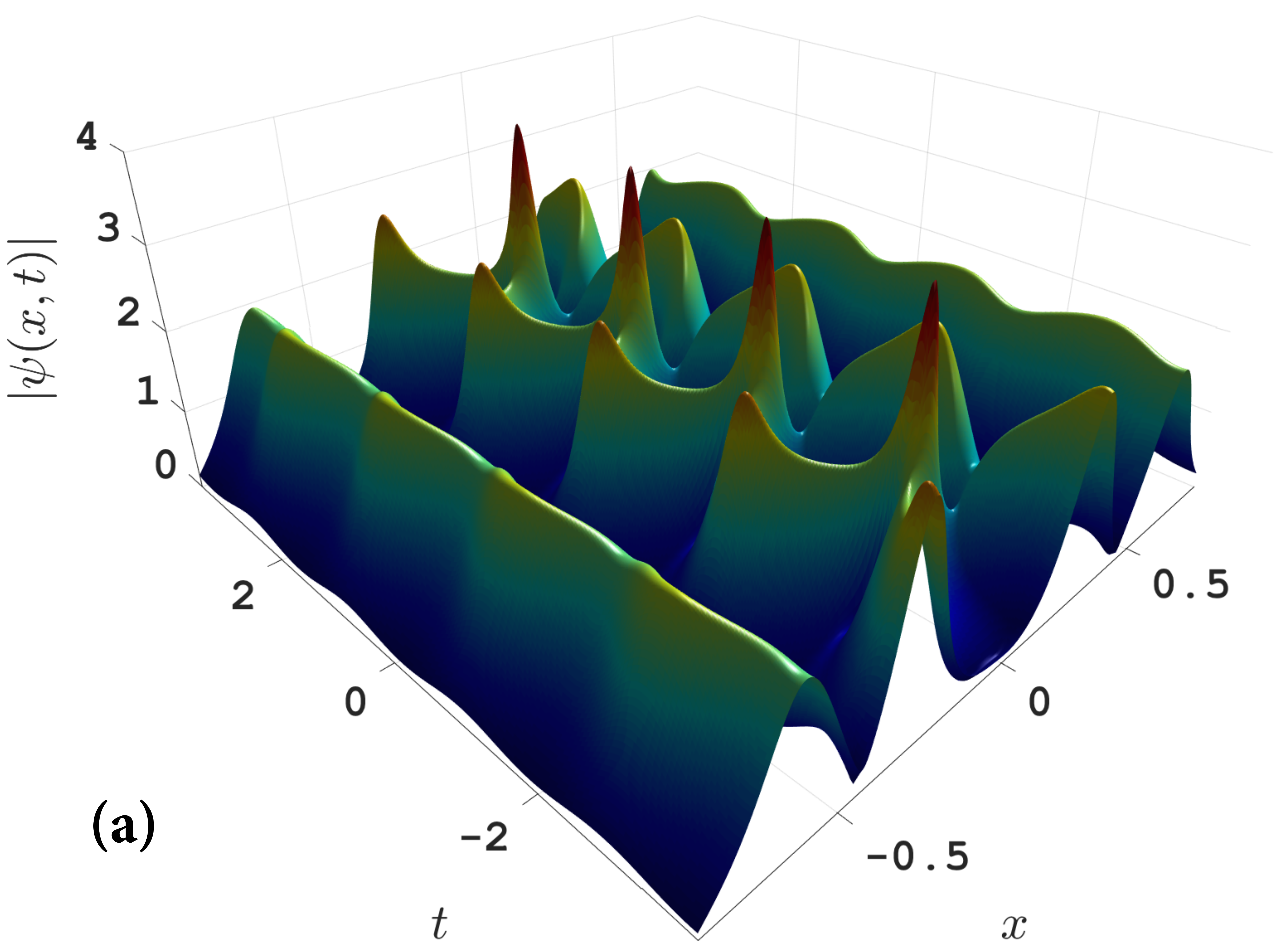}  \quad  \includegraphics[height=1.7in]{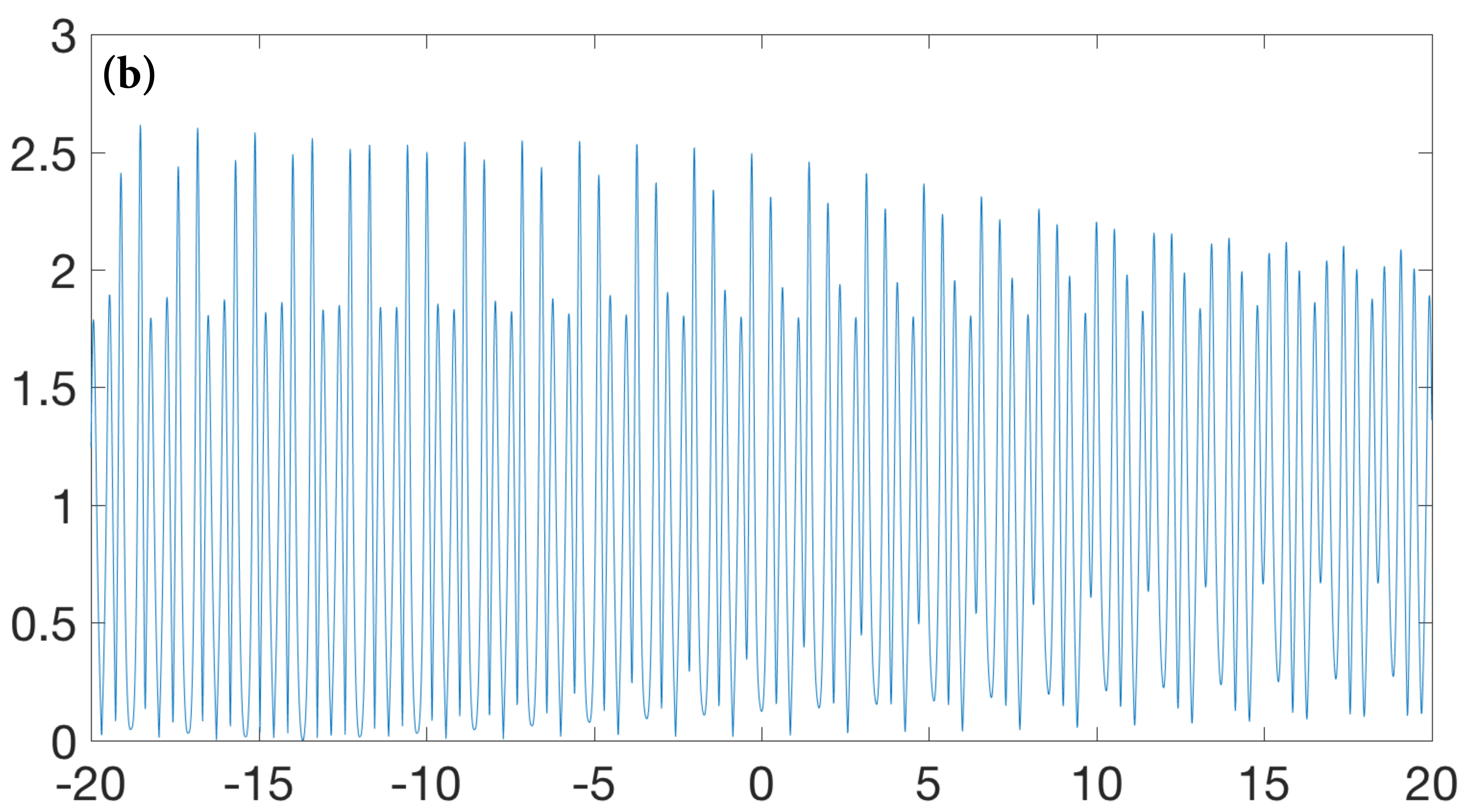}  
 \caption{$g=3$; 
 $\bs{\a} = (-0.28808+i,    -0.09886 +i,   0.096952 +i,     0.28999+i)$;
 $\left\langle |\psi_3|^2\right \rangle=1.4563$;
$K_3=10.9865 $. a) Plot of $|\psi_3(x,t)|$; b) Plot $|\psi_3(x)|$ for a fixed $t$. }
 \label{fig4}
 \end{figure}
\end{center}

\begin{center}
\begin{figure}[ht]
 \includegraphics[height=2in]{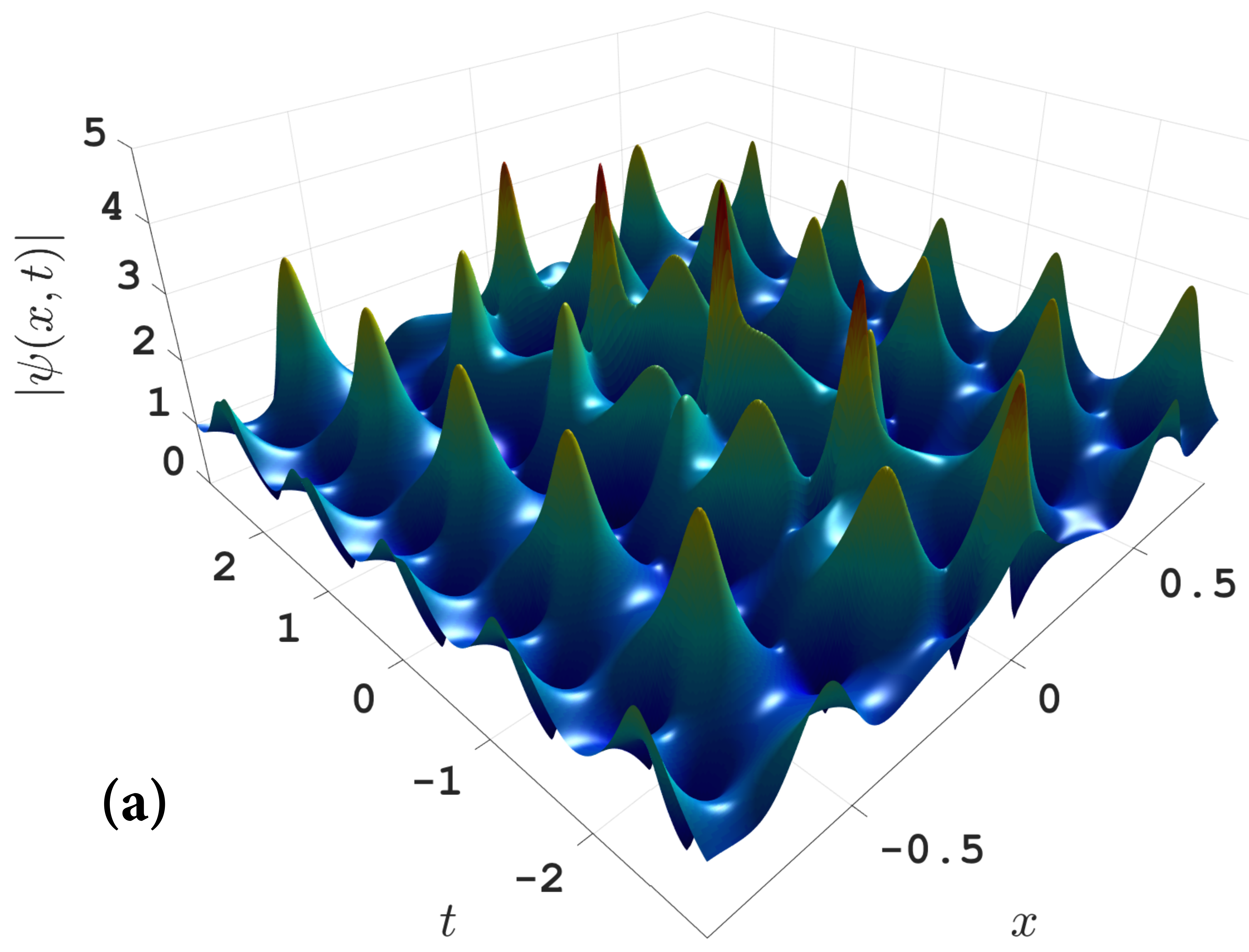}    \quad \includegraphics[height=1.7in]{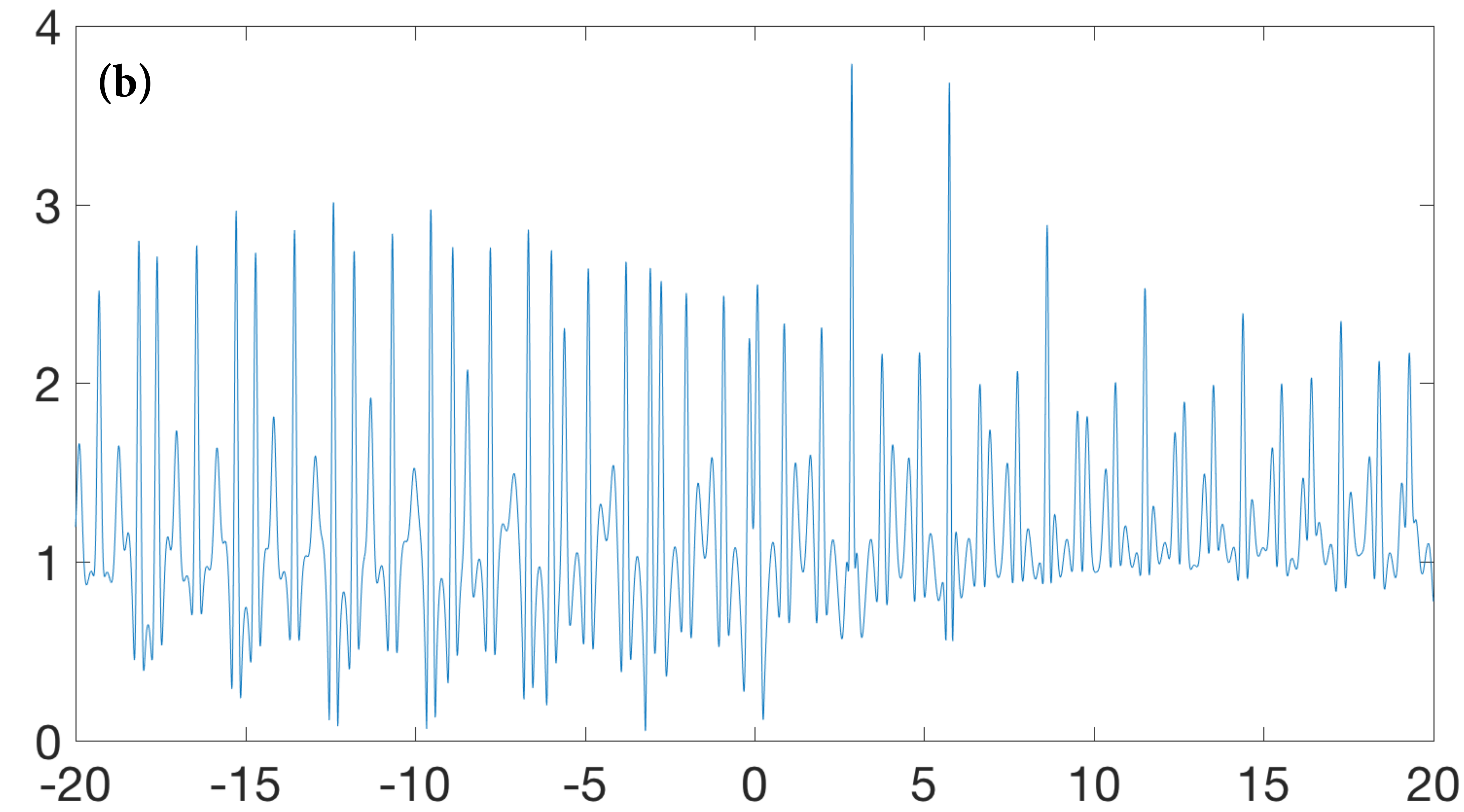}  
 \caption{$g=4$; 
 $\bs\a=(-0.39271 +i,   -0.21336 +i,   0.010556+i,     0.20525+i  ,   0.39027+i)$;
 $\left\langle |\psi_4|^2\right \rangle=1.6452$;
$K_4=15.1959 $. a) Plot of $|\psi_4 (x,t)|$; b) Plot $|\psi_4(x)|$ for a fixed $t$. }
 \label{fig5}
 \end{figure}
\end{center}

 An analogous construction of the finite-band rogue wave identification can be realized for $g>2$, although obviously, the winding of $\mathbb{T}^g$ cannot be as easily illustrated.  In Figures \ref{fig4} and \ref{fig5} we present 3D ($|\psi(x,t)|$) and 2D ($|\psi(x, t^*)|$) plots of the fNLS finite-band solutions  for $g=3$ and $g=4$ respectively. The parameters of both potentials are chosen in such a way that they exhibit rogue waves according to  criterion (\ref{rwc_g}).

\section{Conclusions and outlook}

In this paper, we extend the traditional analytical approach to modelling rogue waves by the special SFB solutions of the fNLS equation.
Our construction uses finite-band ($g$-phase nonlinear wave, $g\in \mathbb{N}$) solutions of fNLS of whom the known SFB solutions (the Peregrine breather, the Kuznetsov-Ma breather and the Akhmediev breather) are particular, degenerate, cases. 
In general, 
a $g$-phase nonlinear wave solution $\psi_g$ is defined by a fixed set of  $g+1$ vertical (spectral) bands  $\g_j\subset\C$ and an arbitrary  vector 
$\bs{\eta}^0$ of $g$ initial phases, where $\bs{\eta}^0 \in \T^g$  - the torus in $\R^g$. More precisely: i) the spectral bands define 
a hyperelliptic Riemann surface $\Rscr$ of genus $g$;  ii)  the finite-band  solution $\psi_g(x,t)$ is expressed in terms of the 
 Riemann Theta functions  \eqref{fingap} on $\Rscr$ so that  $|\psi_g(x,t)| = f_g(\bs{\eta}(x,t))$, where  $\bs{\eta}(x,t)=\bs{k} x+\bs{\o} t +\bs{\eta}^0$ with $\bs{\eta}^0  \in \T^g$.  We introduce the notion of a generalised rogue wave appearing within a $g$-phase nonlinear wave solution $\psi_g$  using the traditional amplitude criterion 
that the maximum of $|\psi_g|^2$ is at least $C_r$ times larger than the mean field $\langle |\psi^{{2}}_g|\rangle$.
The particular value of $C_r$ is at our disposal. To be specific, we use the value  $C_r=8$ suggested by the previous rogue wave studies.
Note that, generically, the wavenumbers $k_j$, $j=1, \dots, g$ are incommensurable, so that by ergodicity the 
mean value  $\langle |\psi^{{2}}_g|\rangle$ over $\mathbb{T}^g$ coincides with mean value of  $|\psi^{{2}}_g(x,t)|$ 
over $x\in \R $ and $\max_{\T^g} [f_g]=\sup_{x\in\R}|\psi_g(x,t)|$  with a fixed arbitrary $t$ in both cases.

The recently obtained \cite{BT} explicit formula for  the maximum of $|\psi_g|$ on $\T^g$ and the formula (\ref{sect-aver})
for $\langle |\psi^{{2}}_g|\rangle$ derived  in the Appendix 
allow us to introduce an explicit criterion \eqref{rwc_g} for the presence of rogue waves within finite-band potentials. Note that this criterion is written in terms
of the Riemann surface $\Rscr$: its branchpoints ${\bs{ \a}, \bs{\bar\a}}$, its normalised holomorphic differentials and some abelian integrals.
Thus, what is essentially described is not a rogue wave, represented by a particular finite-band solution $\psi_g(x,t)$ to the fNLS \eqref{NLS},
but a {\it possibility} of a rogue wave appearance in the family of the finite-band solutions specified by the spectral branchpoints $\bs {\a}, \bs {\bar \a}$ and parametrised by $\bs{\eta}^0\in\T^g$.
Our construction thus opens up a consistent way to introduce a statistical description of the generalised rogue waves 
through the  uniform distribution of the initial phase vector $\bs{\eta}^0$  in $\T^g$. Such a description is currently under the development.

\appendix
\numberwithin{equation}{section}
\section{Computation of the mean intensity}\label{sect-aver}

Here we derive the explicit expression for the average value  of $|\psi^2(x,t)|$ over $x\in\R$. 
Let $\eta(z)$ be  the normalized Abelian integral of the second kind with simple poles at $\infty_\pm$ that has expansions
\be\label{differ-l}
\eta(z)=\pm\le(z+ R_0 + R_1 z^{-1} + O(z^{-2})\ri)~~~~{\rm  so~ that}~~~~~\eta'(z)=\pm\le(1- R_1 z^{-2} +  O(z^{-3})\ri)
\ee
as $z\ra\infty$, where $R_0,R_1\in \C$.
It follows from eq. (6.13) of \cite{IK} that
\be\label{psi-av-init}
\langle |\psi^2| \rangle = 2R_1.
\ee
If  $R(z)=\prod_{j=0}^g(z-\a_j)^\hf(z-\bar\a_j)^\hf$ then direct calculations yield
\be\label{1/R-exp}
\frac{1}{R(z)}=z^{-g-1}\le(1+\frac{\sum_{j=0}^g a_j}{z}+ \frac{(\sum_{j=0}^g a_j)^2+\sum_{j=0}^g (a_j^2-b_j^2)}{2z^2}+ O(z^{-3})\ri).
\ee
Representing the normalized meromorphic differential $\eta'dz$ near $\infty_+$ and taking into account \eqref{1/R-exp}, we obtain
\be\label{eta'}
\eta'(z)=\frac{z^{g+1}-z^g\sum_{j=0}^g a_j+\sum_{j=1}^g \b_jp_j(z)}{R(z)},
\ee
where $\b_j\in\C$ and  
$$
p_j(z)=\k_{j,1}z^{g-1} +  \k_{j,2}z^{g-2} +\dots + \k_{j,g},  \quad j=1,\dots,g
$$ are polynomials  such that 
$\nu_j=\frac{p_j(z)dz}{R(z)}$, $j=1,\dots,g$ are the normalised holomorphic  differentials of the hyperelliptic Riemann surface $\Rscr$ of $R(z)$ with the coefficients 
$\k_{j,i}$ defined by 
\be\label{norm1}
\oint_{\mathbb{A}_k}\nu_j=\d_{k,j}, \ee
where $\d_{k,j}$  is the  Kronecker symbol (see also \eqref{holonorm}).

The normalisation condition of $\eta$ means that $\oint_{\mathbb{A}_k}\eta'(z)dz=0$ for all $k=1,\dots,g$. 
Thus  we obtain
\be\label{beta}
\b_k=-\oint_{\mathbb{A}_k}\frac{z^{g+1}-z^g\sum_{j=0}^g a_j}{R(z)}dz,~~~~~~~k=1,\dots,g.
\ee

Note that according to \eqref{differ-l}, \eqref{eta'}, $-R_1$ is the $O(z^{-2})$ term in the expansion \eqref{eta'} of  $\eta'$ near $z=\infty$.
So, we need to find the leading coefficients $\k_{j,i}$ of the polynomilas $p_j(z)=\k_{j,1}z^{g-1} +  \k_{j,2}z^{g-2} +\dots + \k_{j,g}$, $j=1,\dots,g$. If 
\begin{equation}
 \mathbf{A} =
 \begin{pmatrix}
  \oint_{\mathbb{A}_1}\frac{z^{g-1}dz}{R(z)} & \oint_{\mathbb{A}_1}\frac{z^{g-2}dz}{R(z)}& \cdots \oint_{\mathbb{A}_1}\frac{dz}{R(z)}\\  
   \oint_{\mathbb{A}_2}\frac{z^{g-1}dz}{R(z)} & \oint_{\mathbb{A}_2}\frac{z^{g-2}dz}{R(z)}& \cdots \oint_{\mathbb{A}_2}\frac{dz}{R(z)}\\
   \cdots& \cdots & \cdots &  \\
    \oint_{\mathbb{A}_g}\frac{z^{g-1}dz}{R(z)} & \oint_{\mathbb{A}_g}\frac{z^{g-2}dz}{R(z)}& \cdots \oint_{\mathbb{A}_g}\frac{dz}{R(z)}
 \end{pmatrix}
  ~~~~{\rm then}~~~~\bs{\k}_j=(\k_{j,1},\dots,\k_{j,g})= {\rm Row}_1(\mathbf{A}^{-1})
 \end{equation}
So, according to \eqref{psi-av-init}-\eqref{beta},
\be\label{psi-av-fin1}
\langle |\psi^2| \rangle= 2 \sum_{j=1}^g\k_{j,1}\oint_{\mathbb{A}_j}\frac{z^{g+1}-z^g\sum_{j=0}^g a_j}{R(z)}dz+ \le(\sum_{j=0}^g a_j\ri)^2+\sum_{j=0}^g (b_j^2-a_j^2).
\ee
In particular, without any loss of generality, we can assume $\sum_{j=0}^g a_j=0$. Then \eqref{psi-av-fin1} simplifies to
(\ref{psi-av-fin-simA}).

\section{SFB limit of $\langle |\psi^2| \rangle$ }\label{sect-SFB-lim}

In this subsection we show that the SFB limit (the limit as $a\ra 0$) of $\langle |\psi^2| \rangle$ for a two-phase finite-band solution $\psi$ with the spectral 
bands $\g_0=[-iq,iq]$, $\g_1=[\bar\a,\a]$ and $\g_2=[-\a,-\bar \a]$, where $\a=a+ib$, coincides with the intensity $q^2$ of the background plane wave.
Indeed, in this case $R(z)=\sqrt{(z^2+q^2)(z^2-\a^2)(z^2-\bar\a^2)}$ is an odd function when $z\not\in\g_0$. In this case the leading coefficients
$\k_{j,1}$ of the normalized holomorphic differentials
\be\label{g=2_hol_dif}
\nu_j(z)=\frac{p_j(z)dz}{R(z)},
\ee
where $p_j(z)=\k_{j,1}z+\k_{j,2}$,
can be defined by the system of equations
\be\label{eq-xi}
\le(\oint_{\gt_1}-\oint_{\gt_2}\ri)\nu_1=1,~~~~\le(\oint_{\gt_1}-\oint_{\gt_2}\ri)\nu_2=-1.
\ee
Deforming the contour of integration on $i\R$ and using the fact that $R(z)$ is odd, we reduce the system \eqref{eq-xi} to
\be\label{for-xi}
\k_{1,1}=-\k_{2,1}=\frac{1}{2\int_{iq}^{i\infty}\frac{zdz}{R(z)}}.
\ee
Then, the first sum in \eqref{psi-av-fin1} becomes
\be\label{first-sum}
\sum_{j=1}^2\k_{j,1}\oint_{\mathbb{A}_j}\frac{z^{3}dz}{R(z)}= \frac{\le(2\int_{iq}^{i\r}+\int_\s\ri)\frac{z^{3}dz}{R(z)}}{2\int_{iq}^{i\infty}\frac{zdz}{R(z)}},
 \ee
 where $\r>b$ and $\s$ is a positively oriented arc of the circle $|z|=\r$ from $\frac \pi 2$ to $\frac {3\pi} 2$.

 Consider now the case when $b>q$ and $a\ra 0$, that is, the KM breather limit. Taking the limit $a\ra 0$ in \eqref{first-sum}
 and using the fact that
 \be\label{model}
 \lim_{\e\ra 0^+}\int_{-i}^i\frac{\phi(z)dz}{\sqrt{z^2-\e^2}}=-2\ln\e \phi(0) +O(1),
 \ee
 where $\phi(z)$ is a smooth function that does not depend on $\e$,
 we obtain  
 \be\label{first-sum-KM}
 \lim_{a\ra 0^+}\sum_{j=1}^2\k_{j,1}\oint_{\mathbb{A}_j}\frac{z^{3}dz}{R(z)}= -b^2.
 \ee
 Here we used the fact that the main contribution in the integrals of the right hand side of \eqref{first-sum} come from a
 neighborhood of $z=ib$. Thus, we have shown that in the KM limit ( $a\ra 0$) of $\psi$ we have
 \be\label{SFB-lim}
 \lim_{a\ra 0^+}\langle |\psi^2| \rangle=q^2.
\ee
 
 To obtain the limit \eqref{SFB-lim} in the case of AB (when $b<q$) breather, we choose $\r=q$ in \eqref{first-sum} and represent
 \be\label{first-sum-AB}
 \int_\s\frac{z^{3}dz}{R(z)}=\int_\s\frac{(z^{3}+zb^2)dz}{R(z)} -2b^2\int_{iq}^{i\infty}\frac{zdz}{R(z)}.
 \ee
 To complete the proof of \eqref{SFB-lim} it remains to notice that $\lim_{a\ra 0}R(z)=\sqrt{z^2+q^2}(z^2+b^2)$ and, thus,  
 \be
 \lim_{a\ra 0^+}\int_\s\frac{(z^{3}+zb^2)dz}{R(z)}=\int_\s\frac{zdz}{\sqrt{z^2+q^2}}.
 \ee
 However, the latter integral is zero since the integrand is odd if the contour $\s$ is deformed into $[-iq,iq]$.
 Thus, we proved  \eqref{SFB-lim} for the AB limit.

 \end{document}